# Qubit Logic Modeling by Electronic Circuits and Electromagnetic Signals

G. A. Kouzaev

Alliance of Technology and Science Specialists of Toronto
2-2075 Meadowbrook Rd.
Burlington, ON L7P 2A5, Canada
Tel. 1(905)3318317, E-mail: g132uenf@hotmail.com

**Abstract.** In the paper an approach is presented allowing to model quantum logic circuits by electronic gates for discrete spatially modulated electromagnetic signals. The designed circuitry is for modeling low dimensional quantum nets of general design (up to 25-30 qubits) and quantum devices based only on superposition effect of their work.

1. **Introduction.** One of the trends of the modern electronics is application of various physical effects for development of the highly effective computers. Quantum logic that allows overcoming some problems previously thought unsolvable is an example of the above [1-8]. Unfortunately design a quantum computer meets a number of physical and technological problems; hence the study of quantum mechanical devices is conducted by means, for example, of computer simulation [5,6].

Physical modeling of quantum devices uses well-known similarity between quantum mechanics and physics of electromagnetic waves [9,10]. In [11-13] the idea was developed further to the first gate of pseudoquantum logic working in parallel regime for spatially modulated signals, which fields are discretely (topologically) changed. An optical processor based on similarity between optical and quantum interference has been reported recently [14]. Thus the possibilities of quantum effects modeling by means of macroelectromagnetism or optics attract the increasing attention because of their keen actuality. In this article the new results on quantum gate modeling by electronic circuits for spatial signal processing are discussed.

2. **Topologically modulated field signals and their processing**. It is well known that quantum logical systems have a very unusual and complex behavior that presents considerable difficulties for physical modeling by means of macrophysics [7,8]. But the logical system framework within which the quantum system is supposed to be used greatly limits such behavior, thus allowing imitating it by other methods. One of the approaches is modeling quantum gates by circuits for topological signal processing [9-13]. Earlier the use of these signals was suggested for creation of classical, quasi-neural and multiplace logic as a result of developing the topological approach to the theory of the electromagnetic boundary problems [9,10].

According to the last electromagnetic field can be described by special spatial characteristic - topological chart, that is the set of separatrixes (special field force lines), equilibrium states - points or areas in space where electric, magnetic or electromagnetic fields equal zero (Fig. 1).

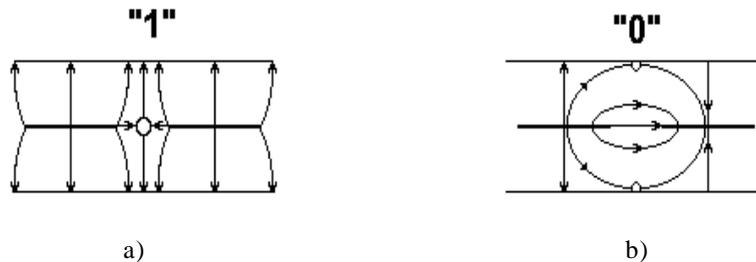

a)                                                      b)
Fig. 1. Electric field force line pictures of even (a) and odd (b) modes of coupled strip transmission lines.

Thus a topological chart acts as a vector-amplitude image of field. Topological chart elements can be derived from differential equations for field force lines; the general theory of such equations was developed by A. A. Andronov and colleagues [15]. In our case field force lines equations look as follows:

$$d\mathbf{r}/dp = \mathbf{E}(\mathbf{r},p), \qquad (1)$$

where $\mathbf{r}$ –radius-vector of a field force line, $\mathbf{E}$-electrical field, $p$-parametrical variable. From (1) it follows that force lines, outlined in space by the end of field vector, behave in non-linear way with respect to summing over fields in the right part of the equation (1). Usually transformations of field force line pictures are quite complicated, but the quality theory of such equations provides an elegant and physically sound approach to their analysis [9,10].

The main idea of this approach is to use field topological charts, as objects with maximum of physical information about fields [9,10]. An interesting feature of topological charts is their structural stability when disturbances do not change the structure or topology of field line pictures. In [9,10] it is shown that in such cases topological charts could be described by a small number of spatial harmonics in a cavity. Outside the area of structural stability topological charts are changed abruptly, as if being the quantum of spatial information (Fig. 2). These changes (bifurcations) can be caused by wave diffraction on waveguide discontinuities or summation of fields. Thus the set of discrete objects - topological charts - can be assigned to the field in waveguide, and the values of logical variables can be assigned to these charts. Then topologically modulated signal can be represented as a sequence of impulses with various field spatial structures $T$. These impulses can also carry discrete information by their amplitudes $A$, so the signal $S(A,T)$ is a two-place object.

The processing these signals can be achieved by various methods. The most common is using predicate logic when predicate is assigned to amplitude $A$, and predicate variable is assigned to topological chart $T$ [16,17]. Then it is possible to design predicate logic gates similar to the known Boolean logic circuits.

According to the second approach the characteristics $A$ and $T$ are considered on equal terms, and multi-valued or multiplace logic is realized [9-13].

In the third method the major element is field topological charts carrying digital information. Passive microwave circuits have been designed allowing to switch topologically modulated signals to different layers of 3D-hybrid integrated circuit, to perform logical operations AND/OR, and also UHF flip-flop circuit [18]. Experimental or theoretical data have been obtained for all developed circuits.

The research of potential limits in discrete electromagnetic image processing has demonstrated the possibility of the creation of subpicosecond passive gates for topologically modulated signals, since the typical time of charge relaxation in conductors does not exceed a few hundredths fractions of a picosecond [9-13,19]. The developed circuits can be used in optoelectronics and high-speed integrated circuits.

So, using 3D-models of electromagnetic signals and electromagnetic theory itself has allowed the modification of signal processing in ICs. The combination of used algorithms can be called electromagnetic logic due to the considerable utilizing of the field effects by the new type of electronic circuits. An interesting expansion of the current approach is an attempt of modeling logic systems based on different physics, quantum mechanics in particular. This possibility exists due to similarity between two phenomena.

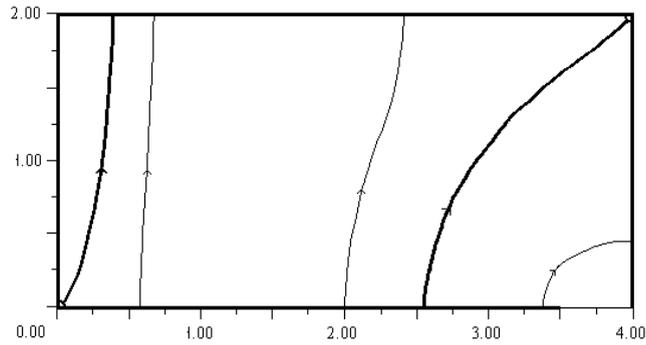

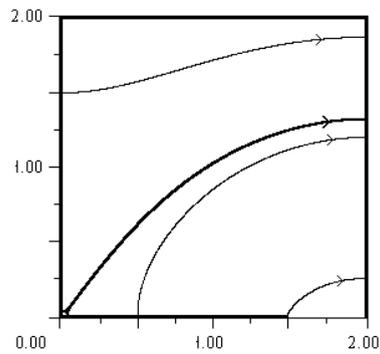

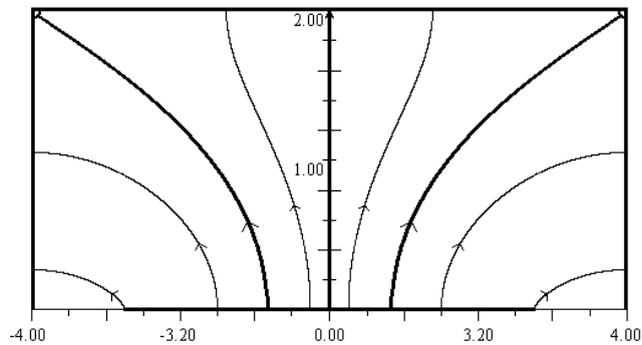

Fig. 2. Electric field force line pictures of excited fields in resonators for different boundary conditions. Exciting frequency $f$ is relatively far from resonance frequency $f_0$. Due to the boundary condition variations topological charts are changed abruptly. The type of the transformation is a separatrix bifurcation caused by changing relationships between spectral components or adding new of them to the fields. The pictures were calculated by using spatial Fourier series and equation (2).

**2. Similarity and difference of quantum and electromagnetic logics.** Let's compare physical and logical characteristics of quantum objects and topologically modulated signals, which are the sequence of field impulses with various electromagnetic field structures. One of the similarity types is caused by analogy between Schredinger equation (2) for probability density $\mathbf{y}$ and dynamical system (3) describing classical movement of a charge in electric field $\mathbf{E}$ :

$$d\mathbf{y}/dt = H\mathbf{y}/(ih/2\mathbf{p}), \qquad (2)$$

$$d\mathbf{r}/dp = \mathbf{E}(\mathbf{r},p), \qquad (3)$$

where $H$ – Hamiltonian, $t$ - time variable, $h$- Plank's constant, $i= \sqrt{(-1)}$, $\mathbf{r}$ – radius vector of a place of a charge. If (2) is a discrete system, then two of its states can correspond to two topologically different sets of charge trajectories describing by (3). So, discrete energy states of an isolated quantum system can correspond to topological charts of electrical field. Electromagnetic field acquires discrete states in waveguides and resonators. The fields of modes have different wave numbers and topological field charts. The number of these states is infinitive similar to the number of energy levels in an isolated quantum system.

Fundamental discrete states of a quantum system may correspond to logical units, for example $/1>$ and $/0>$. The superposition of fundamental quantum states forms a third logical level $(/1> \pm /0>)$ that allows utilizing the physical parallelism of information processing. Waveguide mode fields have topologically varied distributions of force lines, differ from each other discretely and can also have logical variable value assigned [9]. As well as a superposition of two electromagnetic modes has other field topological structure [9]. It can have a new logical variable value assigned and the parallel method of processing the information carried in amplitudes and topological structures of mode fields can be performed [9-13].

Thus the similarity between energy characteristic of quantum system and vector distribution of wave-guide or resonator mode fields is revealed. It allows modeling some of quantum effects by electromagnetic means of the class described above, providing that certain limitations of the analogy are taken into account. One of the limitations is a problem of modeling multiparticle entanglement caused parallelism of quantum computation. In this case modeling quantum circuits by classical signals leads to exponential growth of gate number and connection transmission lines, if a number of qubits is growing linearly.

**3. Qubit modeling theory.** Let's consider shortly some theoretical aspects of qubit modeling. This state can be modeled only approximately from the physics point of view. A qubit should be related to a certain object, which is the product of the superposition of a number of signals and which correlates to the new logical level. The regular signal used in electronics (voltage impulse, for example) loses its component information during the summation of two levels of voltage. In the case of topologically modulated signals the superposition of two signals may cause non-linear transformations of vector-amplitude field image, but each mode building up this image will be still presented in combined signal, and it's processing can produce required parallelism [9]. Lets consider two signals that carry information by their amplitudes and topological field charts: $S_1=(A_1, T_1)$ and $S_0=(A_0, T_0)$, correlating to logical "$1$" and "$0$" accordingly. Superposition of two impulses leads to the summation of their fields and field force line pictures are possible to derive from equation (4):

$$d\mathbf{r}/dp = \mathbf{E}_1(\mathbf{r},p) + \mathbf{E}_0(\mathbf{r},p). \qquad (4)$$

The equation (4) is unlinear regarding to radius-vector *r* of field force lines, and topological chart of combined field $T_{10}$ can be qualitatively different from the charts of the item fields (more detailed information on unlinear transformation of field images is in [9-13]). Mathematically it is expressed by non-homomorphism of these topological objects [9,19]:

$T_{1,0}$-/->$T_1$-/->$T_0$ . (5)

An example of the effect is in Fig.2, where different forms of exiting hole in a resonator cause different spectral mixture of spatial harmonics and topologically different pictures of field force lines.
Thus the new topological chart can be assigned with the new value of logical variable. Formally this feature of topologically modulated signals resembles the feature of qubit, and, since the initial signals are still present in the combined field, which processing as a total will cause parallel processing of these signals, it can be used for modeling certain effects in quantum circuits. Methods and approaches of electronics and techniques of microwave electromagnetic signal processing should be used for the creation of modeling circuits. In present a number of such devices has been developed that can be used as prototypes for pseudo quantum logic circuits.

5. **Electronic gates for qubit logic modeling.** General principle of quantum calculation consists of fulfillment of physical operations with N-set of quantum particles. The operations are possible to describe with a logical net. Number of logical operations due to quantum parallelism is $2^N$. Macroelectromagnetism allows realizing the operations only in real space, hence the number of required gates should grow in exponential manner. Behind of that, the electronic modeling net with dimension no more than several tens qubits can be useful for studying quantum logic, development of control pseudoquantum circuitry in solid-state quantum computers and etc. Let's consider possible gates for pseudoquantum signal processing. The circuitry depends on a type of topologically modulated signal using for quantum logic modeling. One set of the signal is showed on Fig. 1. Logical "*0*" and "*1*" correspond to odd and even modes. Superposition of fields represents exited field near only one conductor of a coupled strip transmission line and it is topologically different from odd and even mode field force line pictures (Fig.3).

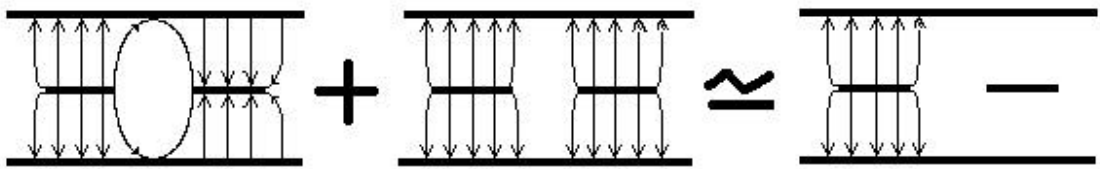

Fig. 3. Superposition state of fundamental modes as a qubit model.

The field transformation is possible to describe by transmission matrix *T*. Elements of *T* are transmission coefficients of a scattering matrix of a device. The *T*-matrix is an analog of operator *U* from quantum mechanics and should be unitary, that provides logical reversibility of operations.

It is well known that quantum logic requires two-bit gates and several one-bit gates [1-7]. Let's consider additional devices that are needed for realizing the gates. First of them is a passive mode filter for transmission of an even mode signal to the output of the device (Fig. 4, a, b). Odd mode is suppressed essentially due to the filter design. The filtration effect is illustrated by Fig. 5, 6, where microwave current distributions along the device are shown. Both distributions have steps near the discontinuity, but only odd mode current is decreased about exponentially after junction between coupled and single strip transmission lines. So, the device is an analog of a diode gate from digital electronics.

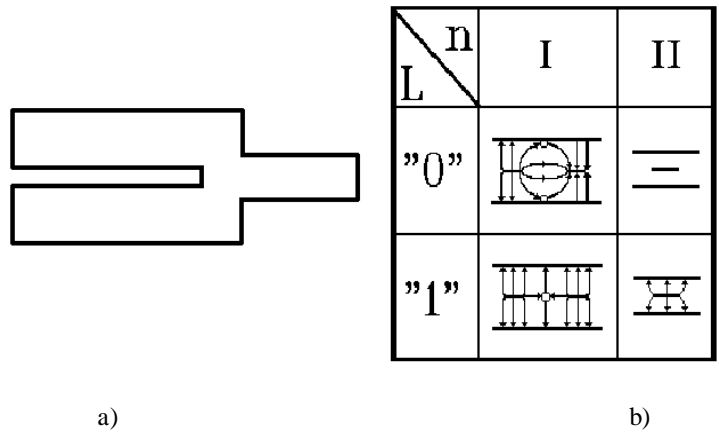

a)                                              b)

Fig. 4. A spatial filter for topologically modulated field signals designed on connection between coupled and single strip transmission lines a). Truth table of the filter b): *n=I*-input, *n=II*- output, *L*- logical levels.

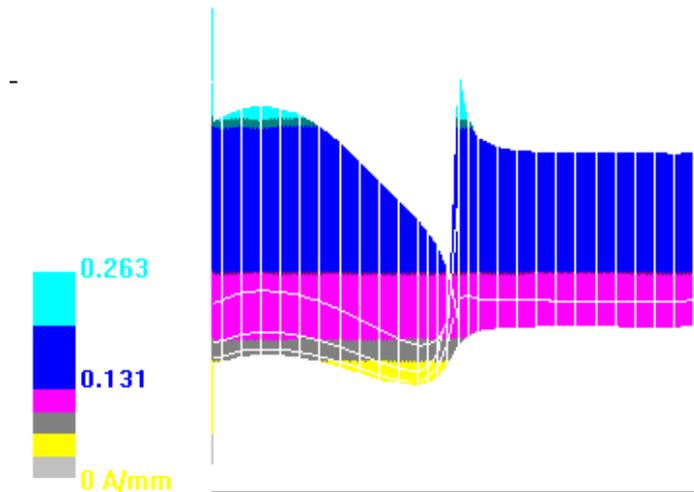

Fig. 5. Microwave current distribution for even mode diffracting on the junction of coupled and single microstrip transmission lines. The circuit is a pass-device for the mode.

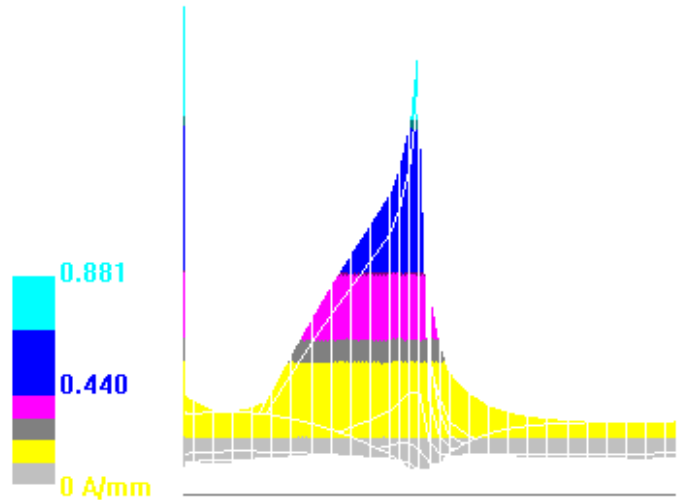

Fig. 6. Microwave current distribution of odd mode along the filter. The circuit suppresses the mode.

Another gate (Fig. 7) named a switch of topologically modulated signals is a reversible demultiplexer/multiplexer theoretically and experimentally studied in [9-13,19]. The gate plays an important role for demultiplexing even and odd modes and can work in parallel manner for superposition state of input signals. Combination of the gate and transistors allows realizing any logical function in digital processing amplitude-spatially modulated impulses of electromagnetic field.

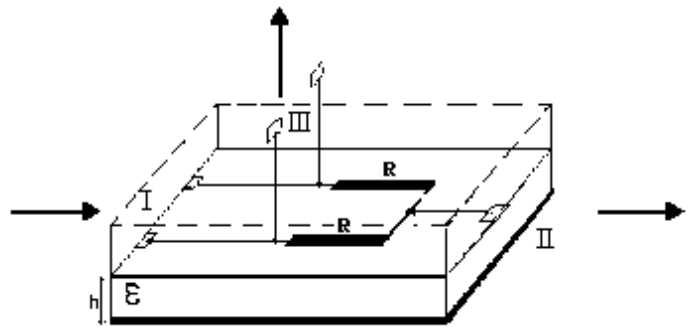

Fig. 7. Switch for topologically modulated impulse field signals. $I$ - input of the signals (coupled strip transmission lines with characteristic impedance $Z_e$ and $Z_o$), $II$ - output of logical "1" (two-conductor line with characteristic impedance $Z^{II}$, $III$ - output of logical "0" (strip transmission line with characteristic impedance $Z^{III}$).

Boolean gates of passive type (NOT, OR, AND, Flip-Flop) have been considered in [9,18-20] and designed on the interference principle. A multivalued switch containing diodes for joint amplitude-spatial signal processing has been studied experimentally and theoretically recently [9-13]. All devices are able to work in parallel manner and pertinent to be prototypes for quantum gates modeling.

The preliminary research allowed to design and experimentally tested circuits capable to model quantum gates [10-13]. Between them is a NOT gate (Fig. 8-10) working in reversible and parallel manner. The gate inverts images of fundamental modes and their superposition states into each other.

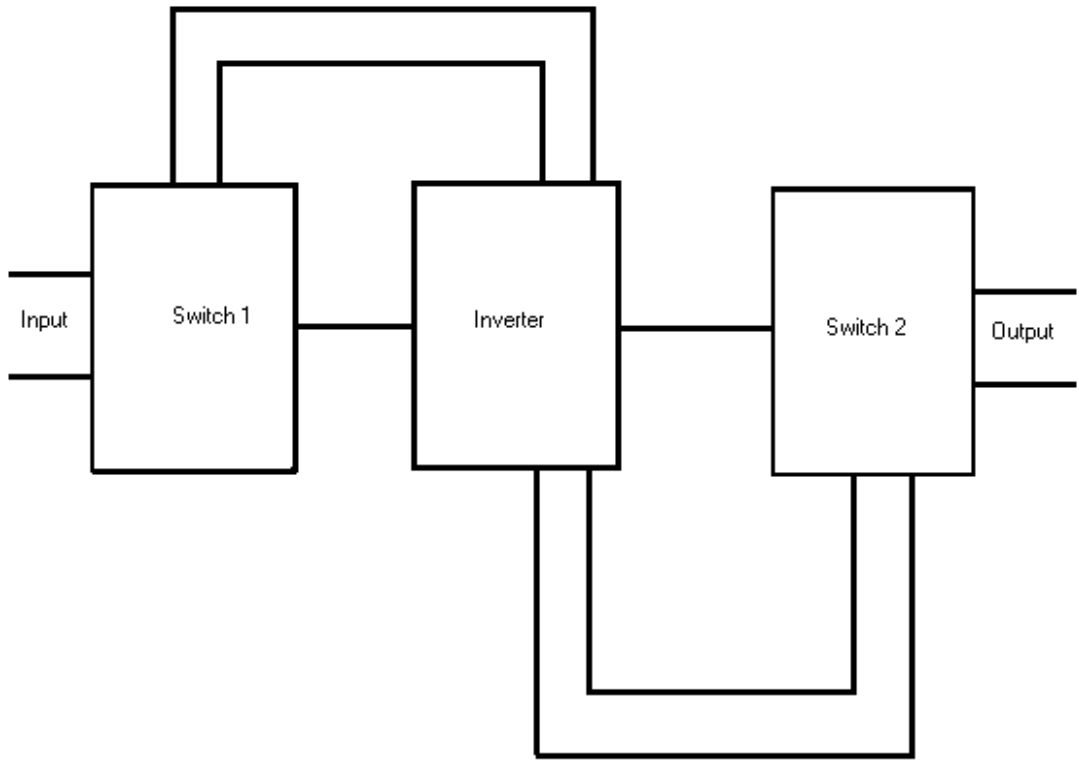

Fig. 8. NOT gate for parallel signal processing and its truth table. *I*-input signal, *II*- output signal.

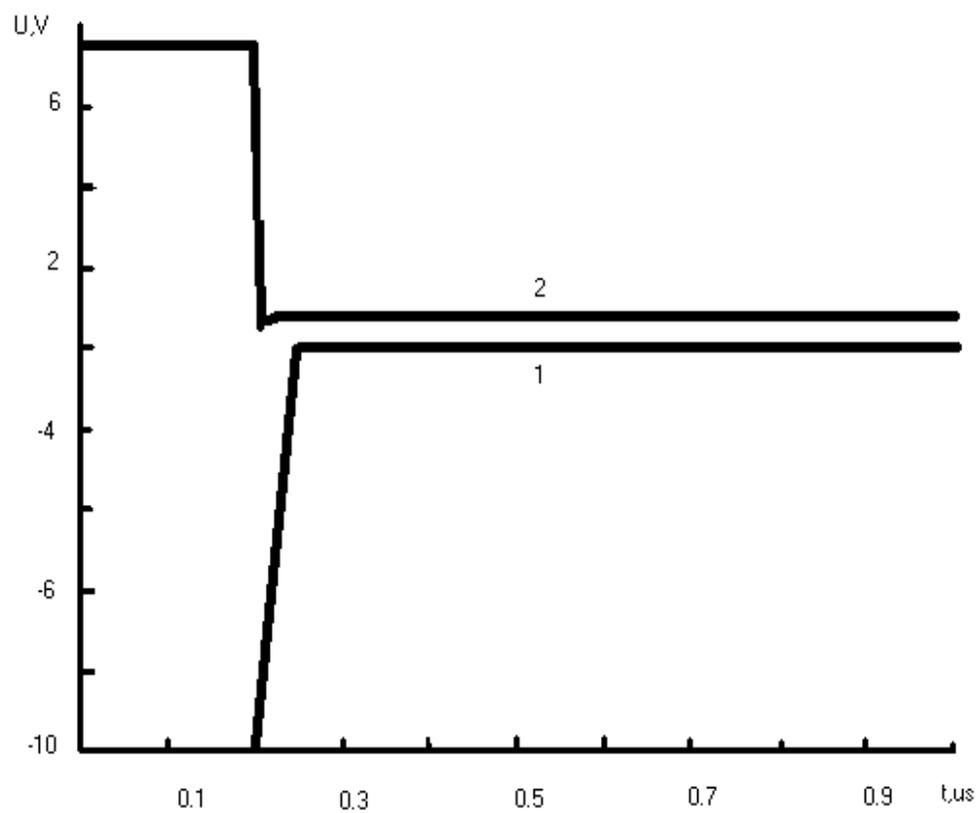

Fig. 9. Results of modeling NOT gate for inverting logical units corresponding to different modes of a coupled strip transmission line. 1- input voltage of even mode of coupled strip transmission line, 2 - output voltage of odd mode of coupled strip transmission line.

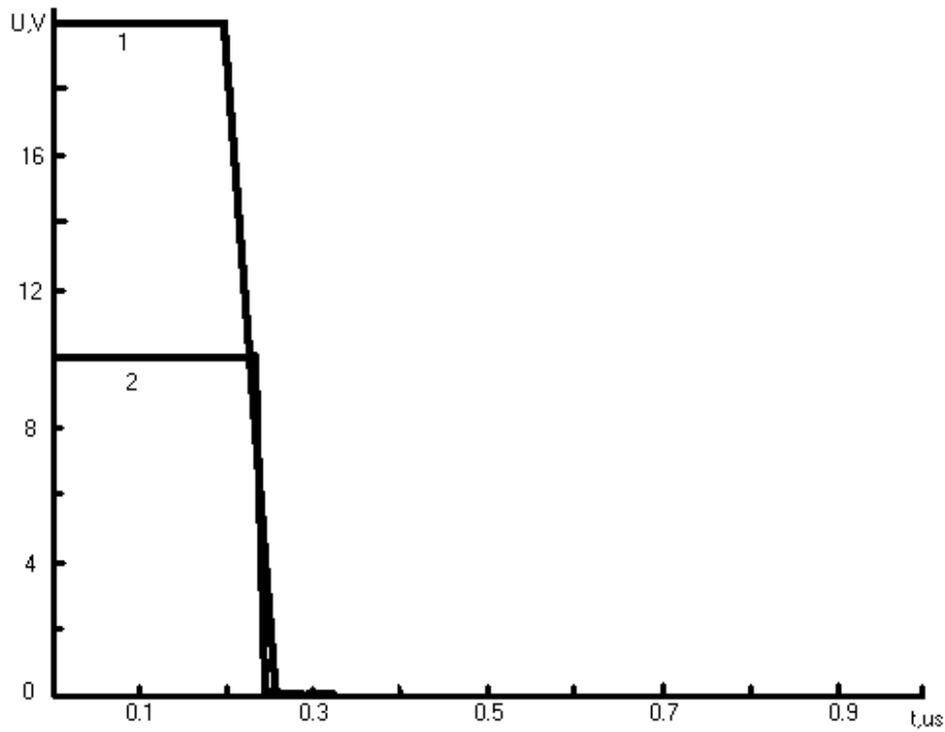

Fig. 10. Results of modeling NOT gate for inverting logical units corresponding to different modes of a coupled strip transmission line. 1- input (differential) voltage of odd mode of coupled strip transmission line; 2- output voltage of even mode of coupled strip transmission line.

Another gate exclusively used in quantum logic is √NOT. The gate transforms fundamental signals (modes) into a mixture of them. Fig. 11 shows truth table of the designed gate for topologically modulated signals.

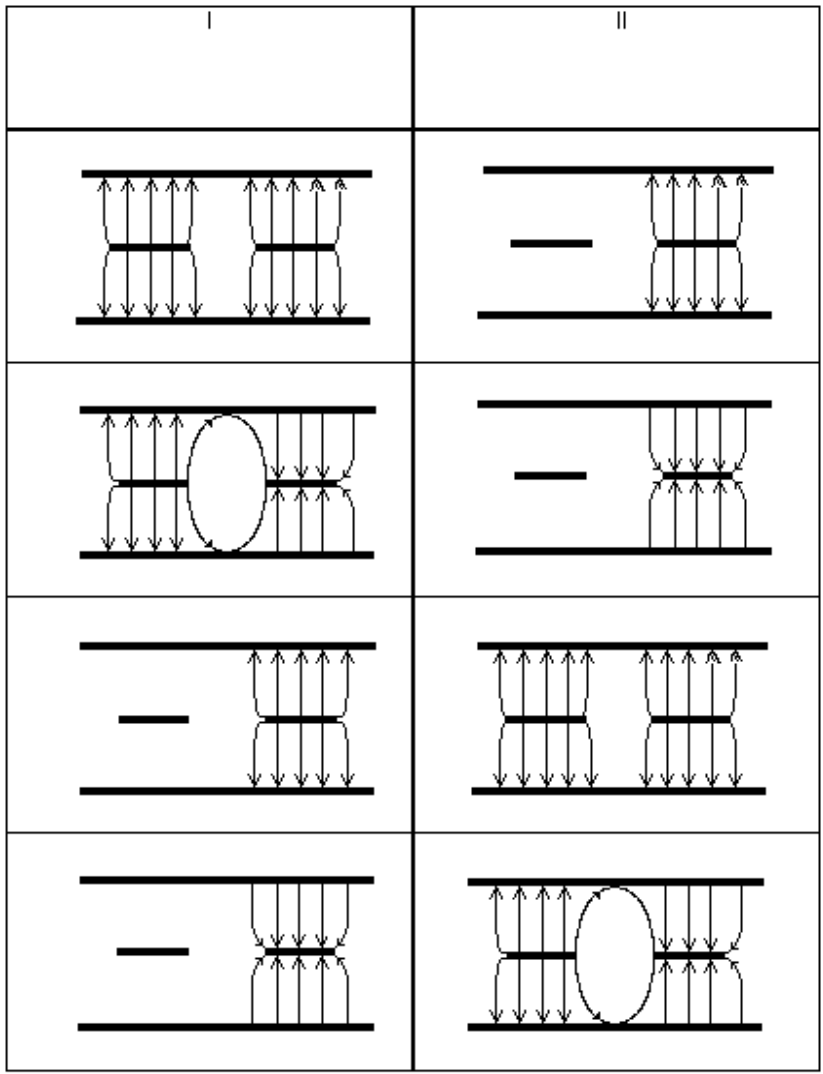

Fig. 11. Truth table of √NOT gate for topologically modulated signals. *I*-input signal, *II* –output signal.

Next important gate realizes controlled NOT (CNOT) operation. The gate was designed on a combination of passive and active components like NOT gate and has 2 inputs and 2 outputs for providing logical reversibility. Field impulses from controlling transmission line stops or stimulates field distribution changing on the output of the device working for single or mixture mode regimes (Fig. 12).

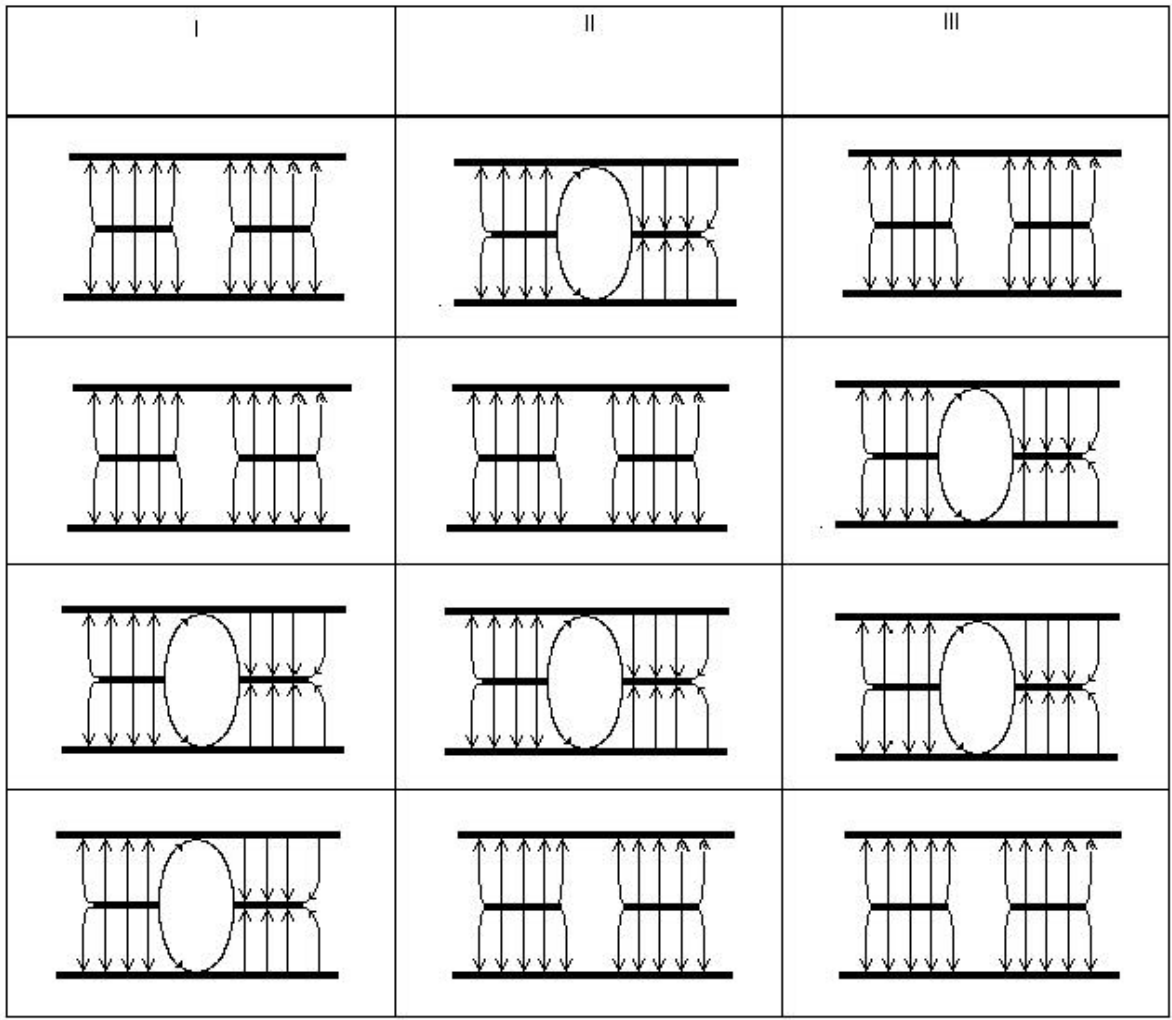

Fig.12. Truth table of CNOT gate for topologically modulated signals. *I*- controlling signal (input/output), *II*-input controlled signal, *III*-controlled output signal.

The designed gates allow modeling quantum nets using combination of the devices. Maximal number of electronic gates is limited technologically and is about 1 billion gates on a chip working with clock frequency about 10-20 gigahertz. So, one chip will be able to model a quantum net consisting of 25-30 qubits with "microwave speed". Certain decreasing required electronic gate number is possible to reach using known Quine-McCluskey algorithm of logical circuit design [21]. Another way is the searching different physical effects for more effective modeling quantum entanglement caused high level of quantum computer parallelism [22].
The developed method of modeling can be used for testing quantum algorithms, for realizing hybrid quantum-conventional computers of solid-state design [23], and for modeling quantum algorithms realized only on the base of superposition effects and etc.

**6. Conclusion.** In the paper an approach and electronic gates are proposed allowing to model qubit logic by classical means. Behind of the exponential growth of required number of electronic gates the circuitry will be able to model quantum nets consisting of about 25-30 qubits. Another field of application of the design principles is quantum-conventional computers of solid-state design and quantum algorithms that require only linear growth of number of logical states.

**Acknowledgements.** The author thanks The Russian Foundation of Basic Research for financial support of the work. I am appreciating the support of Moscow State Institute of Electronics and Mathematics (Technical University) where I did the research. As well as the author is grateful to Gennum Corp. (Canada) for giving a time to finish the paper preparation.